\documentstyle[emulateapj,psfig]{article}
\begin{document}

\newcommand{\msun}{$M_{\odot}$}
\newcommand{\etal}{{\it et al. }}
\submitted{Accepted for publication in ApJ, 2001, Vol. 552} 

\title{
Testing the transition layer model of quasi-periodic oscillations in 
neutron star
X-ray binaries
  }
%
%
\author{Xue-Bing Wu$^{1,2}$}
%
%
%
\affil{$^1$
       Beijing Astronomical Observatory, Chinese Academy of
       Sciences, Beijing 100012, China\\
	$^2$
	Department of Astronomy, Peking University and CAS-PKU Joint Beijing
Astrophysics Center,  Beijing 100871,        China \\
wuxb@bao.ac.cn}

\begin{abstract}

We compare the theoretical predictions of the transition layer model with some
observational features of quasi-periodic oscillations (QPOs) in
neutron star X-ray binaries. We found that the correlation between 
horizontal branch oscillation (HBO) frequencies and
kilohertz (kHz) QPO frequencies, the difference between the low-frequency QPOs
in atoll sources and HBOs in Z sources, and the correlation between the
frequencies of low-frequency QPOs and break frequencies can be well
explained by the transition layer model, provided the neutron star
mass is around 1.4 solar mass and the angle between magnetosphere equator and
accretion disk plane is around 6 degree. The observed decrease of peak
separation between two kHz QPO frequencies with the increase of 
kHz QPO frequencies and the increase of QPO frequencies
with the increase of inferred mass accretion rate  are also
consistent with the theoretical predictions of transition layer model. 
In addition, we derive a simple equation that can be adopted to 
estimate the angle ($\delta$) between magnetosphere equator and accretion disk
plane by use of the simultaneously observed QPO frequency data. We
estimate these angles, in the range of 4 to 8 degrees,  for five Z
sources and two atoll sources. The nearly constant $\delta$ value for
each source, derived from the different sets of simultaneously
observed QPO frequency data, provides a strong test of the theoretical
model.  Finally, 
we suggest that the similar 
transition layer oscillations may be also responsible for the observed QPOs in
accretion-powered millisecond X-ray pulsar and Galactic black hole candidates.

\end{abstract}


\section{Introduction}
Quasi-periodic Oscillations (QPOs) have been found to be a very common feature 
of accreting systems around compact objects. In neutron star low-mass X-ray 
binaries (LMXBs) with low magnetic field, at least four types of QPOs,
namely the 15-60 Hz horizontal branch oscillations (HBOs; van der Klis
et al. 1985), the 6-20 Hz normal branch oscillations (NBOs; Middledich
\& Priedhorsky 1986) and the 300-1200 Hz lower and upper kilohertz
(kHz) QPOs (van der Klis et al. 1996), have been 
discovered by several X-ray satellites including {\it EXOSAT, GINGA}, and the {\it Rossi X-Ray Timing Explorer
(RXTE)}. In spite of the different evolution tracks in the X-ray color-color 
diagram (Hasinger \& van der Klis 1989), the low-frequency QPOs in
atoll sources seem to have the similar feature to those of the HBOs in Z
sources (Strohmayer et al. 1996, Wijnands \& van der Klis 1997). With
the advantage of {\it RXTE}, 20 sources 
have now shown kHz QPOs and 18 of them have shown two simultaneous kHz peaks 
(see van der Klis 2000 for a review).
The peak separations of two kHz QPOs have been observed to decrease
considerably in five sources 
when the kHz QPO frequencies increase (van der Klis \etal 1997;
M\'endez et al. 1998a,b; 
Ford et al. 1998, M\'endez \& van der Klis 1999, Markwardt et al. 1999). Other 
sources may also show similar variations (Psaltis et al. 1998). It has been 
found that the HBO frequencies of Z sources and the frequencies of 
low-frequency QPOs in atoll sources 
are well correlated with the kHz QPO frequencies (Psaltis et
al. 1999a). Similar correlation of low and high frequency QPOs might
also  exist for some other neutron star systems and Galactic black hole
candidates. In addition, 
tight correlations of the HBO 
frequencies of Z sources and of the frequencies of low-frequency QPOs in
atoll sources and black hole candidates with the break frequencies
shown in their power density 
spectra have also been found recently (Wijnands \& van der Klis
1999). These two 
correlations strongly imply that similar mechanisms may be responsible for 
the break frequency, the low-frequency QPOs and the high-frequency QPOs in
both neutron star and black hole X-ray binaries.

There are several QPO models that are currently in debate. Due to the 
difficulty in producing kHz QPOs, the previous magnetospheric beat-frequency 
model proposed by Alpar \& Shaham (1985) has been modified as
sonic-point beat-frequency model (Miller, Lamb \& Psaltis 1998). In this
model the upper kHz QPOs 
arise from the clumps moving with the Keplerian frequency at the sonic point 
that is near the innermost stable circular orbit of neutron star. The
lower kHz QPO frequency is thus the beat frequency between the upper
kHz QPO frequency and 
the neutron star spin frequency. However, this model, if without
further modifications, predicts a
constant peak separation between two kHz QPO frequencies, which is
contrary to the 
observations.  Moreover,
recent investigation on the boundary layer accretion onto neutron star
indicated that the accretion flow is always subsonic and there is
probably no such a sonic point at all near the neutron
star surface (Popham \& Sunyaev 2001). This leads to further doubt
about the sonic-point
beat-frequency model.
Another attractive model is the  relativistic precession 
model(RPM) proposed by Stella \& Vietri (1998, 1999), who identified the upper 
kHz QPO frequency with that of an Keplerian orbit in the disk and the
lower kHz 
QPO frequency and HBO frequency with, respectively, the periastron
precession frequency and 
twice of the nodal precession frequency of this orbit. This model can qualitatively 
explain the observed correlations between these three QPO frequencies
(Stella, Vietri \& Morskink 
1999), but the precise match requires some fine tuning of additional free 
parameters such as the orbital eccentricity and the periastron
distance.  Moreover, in 
order to explain the HBOs and its correlation with kHz QPOs, this model requires 
that the $I/M$ value (where $I$ is the momentum of inertial and $M$ is the
mass of neutron 
star) be 2 times larger than that predicted by realistic neutron star 
equation of state (Psaltis et al. 1999b). In addition, RPM model
requires the neutron star 
mass in the range of 1.9-2.2\msun~in order to explain the correlation
between low and high QPO frequencies, which is significantly larger than the 
measured value, 1.1\msun$<$M$<$1.6\msun, for neutron stars in radio
pulsar binaries (Thorsett \& Chakrabarty 1999; Finn 1994). The problem to make a 
neutron star with mass larger than 2.2\msun~arises because it would require accretion 
of 
material of at least 0.8\msun, which means that these X-ray binaries would have to be 
rather old and would have to have spun up rapidly for 90\% of their lifetime (Lai, Lovelace \& 
Wasserman 1999). 
The same problem probably exists for the new model proposed by
Psaltis \& Norman (2000), who derived the similar characteristic QPO frequencies
as in RPM in the inner accretion disk.

Another alternative model for QPOs in neutron star X-ray binaries is the disk 
transition layer model  (hearafter TLM) proposed recently by Titarchuk \& 
Osherovich (1999), Osherovich \& Titarchuk (1999a,b) and
 Titarchuk, Oshervich \&
Kuznetsov (1999). The geometry of the transition
layer model has been clearly shown in  Figure 1 of Titarchuk 
\etal (1999). 
In this ``two-oscillator'' model, the lower kHz QPO frequency is 
identified with the Keplerian frequency 
in a viscous 
transition layer between Keplerian disk and neutron star surface. 
 Assuming the magnetic field of neutron star is low (in the range
of $10^7$ to $10^9~ Gauss$), the Alfv\'en radius, depending mainly on the
magnetic field and accretion rate,
is at most  of several radii of neutron star. The size of
the magnetosphere, as well as that of the transition layer, is thus very
small.
Viscous oscillations in this transition layer produce the observed noise break
and a low frequency QPO. 
 In addition, the adjustment of the Keplerian disk to the 
sub-Keplerian layer may create conditions favorable for the
formation of  a hot blob in the transition layer.
This blob, when thrown out into the rotating magnetosphere from the
transition layer, participates in the radial oscillations with Keplerian
frequency.  Under the influence of the Coriolis force,
such a blob, assumed to be a Keplerian oscillator, 
oscillates both radially and perpendicular to the disk.
This produces
two harmonics of another low-frequency QPO (HBO in Z sources) and the
upper kHz QPO. These six different frequencies have been identified in
two atoll sources (4U 1728-24 and 4U 1702-42) and Z source Sco X-1 (
Titarchuk et al. 1999; Osherovich \& Titarchuk 1999a,b). The
observed correlations of QPO frequencies for these sources seem to be 
consistent with the
theory of TLM.

In this paper, we derive more theoretical predictions for TLM and
compare them with the available observational
data of QPOs in neutron star X-ray binaries. The consistency between
model predictions and observational data without assuming larger
neutron star mass and a stiff equation of state suggests that TLM is
perhaps a more competitive model than others. We also suggest that a
similar model may be also applied to the cases of millisecond X-ray
pulsars and Galactic black hole candidates.

\section{Characteristic QPO frequencies in the transition layer model}
As indicated by Titarchuk \& Osherovich (1999) and Osherovich \&
Titarchuk (1999b), there are six
characteristic QPO frequencies in the TLM. First, the fundamental
frequency is  assumed to be the Keplerian frequency, 
namely,
\begin{equation}
\nu_k=\frac{1}{2\pi}(\frac{GM}{R^3})^{1/2},
\end{equation}
where $G$ is the gravitational constant, $M$ is the mass of neutron
star and $R$ is the radius of  an orbit in the transition
layer. The linear Keplerian oscillator with frequency $\Omega_k/2\pi$ in the frame of reference
rotating with rotational frequency $\Omega/2\pi$ (not perpendicular to the
disk plane) is known to have an exact
solution describing two branches of oscillations (see \S 39 in Landau \& lifshitz
1960). Assuming $\Omega$ is constant, the dispersion relation of the
frequency of oscillation $\omega$ can be derived as:
\begin{equation}
\omega^4-(\Omega_k^2+4\Omega^2)\omega^2+4\Omega^2\Omega_k^2 sin^2\delta=0,
\end{equation}
where $\delta$ is the angle between $\Omega$ and the normal to the
Keplerian oscillation. If $\delta$ is assumed to be small, one can
obtain simple expressions of two characteristic oscillation
frequencies from the dispersion relation.
The radial eigenmode has a frequency
\begin{equation}
\nu_h=\sqrt{\nu_k^2+(\frac{\Omega}{\pi})^2},
\end{equation}
where $\Omega$ is the angular
velocity of the rotating magnetosphere in the case of neutron star
X-ray 
binaries. If the magnetosphere corotates with the neutron star, we
have $\Omega=\Omega_0$ where $\Omega_0$ is the rotating angular velocity of
neutron star. 
The frequency $\nu_h$ is the analog of the hybrid frequency in
plasma physics (Akhiezer \etal 1975; Benson 1977). On the other hand, 
the frequency of the vertical eigenmode is
\begin{equation}
\nu_L=\frac{\Omega}{\pi}\frac{\nu_k}{\nu_h}sin\delta,
\end{equation}
 These three
characteristic frequencies, $\nu_k$, $\nu_h$ and $\nu_L$,  are assumed
to account for the frequencies of lower kHz QPOs, upper kHz QPOs and HBOs
in Z sources, 
respectively. Because the second harmonics of HBO are often seen in Z
sources, they are interpreted as 
 $2\nu_L$, counting as the other characteristic frequency. 

From equations (3) and (4), we can derive several very useful
relations among these characteristic frequencies. First, the
correlation
 between the HBO
frequency and two kHz QPO frequencies can be expressed as:
\begin{equation}
\nu_L=\nu_k\sqrt{1-(\frac{\nu_k}{\nu_h})^2}sin\delta.
\end{equation}
Second, the difference between two kHz QPO frequencies can be related
to $\nu_k$ as
\begin{equation}
\Delta\nu=\nu_h-\nu_k=\nu_k(\sqrt{1+(\frac{\Omega}{\pi\nu_k})^2}-1).
\end{equation}
Third, if the Keplerian oscillation is aligned with the accretion disk plane, 
the angle between magnetosphere equator and disk plane can be
determined by
\begin{equation}
\delta=arcsin(\frac{\nu_h\nu_L}{\nu_k\sqrt{\nu_h^2-\nu_k^2}}).
\end{equation}
Because three frequencies in the right of equation (7) are all
observational quantities, this equation  perhaps suggests a useful
method to
determine $\delta$ for neutron star X-ray binaries. 

The other two characteristic frequencies in the TLM are related to the two
typical timescales in the transition layer. One is the radial drift
timescale of matter through the layer bounded between disk and
neutron star, namely, $t_r\sim(R_{out}-R_0)/v_r$, where $R_0$ is the
radius of neutron star (taken to be $6GM/c^2$ in this paper) and $v_r$ is the 
radial drift velocity. This timescale gives the characteristic 
 viscous frequency as
\begin{equation}
\nu_v\simeq \frac{\gamma \nu}{(R_{out}-R_0)R}\simeq 2\pi\alpha\gamma\nu_k
(\frac{H}{R})^2\frac{r_{out}}{r_{out}-1},
\end{equation}
where $r_{out}$ is defined as $r_{out}=R_{out}/R_0$, and $\gamma$ is the 
Reynolds number given by
$\gamma=\dot{M}/4\pi\rho\nu H=v_r R/\nu$ where
$\dot{M}$ is the mass accretion rate and $\rho$ is the mass density. We adopt 
the
standard viscosity prescription $\nu=\alpha c_s H=\alpha\Omega_k H^2$,
where $c_s$ is
the local sound speed and $H$ is the scale height (Shakura \& Sunyaev
1973). The relation
between $\gamma$ and $r_{out}$ can be obtained by considering the radial
transport of angular momentum and the boundary conditions at $R_0$ and
$R_{out}$ (Titarchuk \& Osherovich 1999).
Assuming that at $R_0$, $\Omega=\Omega_0$ 
and at $R_{out}$,
$\Omega=\Omega_{k}$ and $d\Omega/dr=d\Omega_{k}/dr$ and defining
$A=\Omega_{k0}/\Omega_0$  (where
$\Omega_{k0}$ 
is the Keplerian angular velocity at $R_0$), 
$r_{out}$ can be determined by following equation:
\begin{equation}
(\frac{3}{2}-\gamma)Ar_{out}^{2-\gamma}+(\gamma-2)r_{out}^{-
\gamma}+\frac{1}{2}A=0.
\end{equation}

Another characteristic timescale in the transition layer is the
diffusion time scale given by $t_{diff}\sim(R_{out}-R_0)^2/ 
(l_{d} v_r)$, where $l_{d}$ is the diffusion length scale in the
transition layer.  
It gives the break frequency as
\begin{equation}
\nu_{break}\simeq\frac{l_{d}}{(R_{out}-R_0)}\nu_v=\frac{l_{d}}{R_0} 
\frac{\nu_v}{r_{out}-1}.
\end{equation}
This expression of break frequency is similar as that given in
equation (29) of Psaltis \& Norman (2000) if $l_{d}$ is replaced by
$R$. 
 Similar to the argument given by  Psaltis \& Norman (2000),
the reason that $\nu_{break}$ observed as a frequency break may be that the 
transition layer acts as a filter band so that the response is
constant if the oscillation frequency is lower than the inverse
diffusion time scale and the response decreases above it.
Titarchuk \& Osherovich (1999) described $l_{d}$ as the mean
free path of a particle, however, this introduced another uncertain
quantity in the TLM theory. In this paper,  using a similar approach
adopted in the standard accretion disk model (Shakura \& Sunyaev
1973), we assume that $l_{d}$ is
smaller than the vertical and radial size of the transition layer and
take ${l_{d}}/{R_0}$ as a constant less than unity for simplicity
(in \S 3.3 
we will show this assumption
is consistent 
with the observed data of break frequency).
The equations (8), (9) and (10) can be used to calculate the correlations
between other low-frequency QPO (besides HBO) frequency,  break
frequency and Keplerian frequency. The original descriptions of viscous and break
frequencies shown in equations (12) and (13) in Titarchuk \&
Osherovich (1999) have been improved here (see equations (8) and (10)) by
considering a more explicit expression of the radial drift timescale.

\section{Comparisons of model predictions with the observational data}
There are a lot of observations made by several X-ray satellites on QPOs in 
neutron star X-ray binaries. But only until very recently the correlations among 
the frequencies of low and high-frequency QPOs, the variation of kHz QPO 
frequencies and the difference between HBOs and other low-frequency QPOs 
have been systematically studied with plenty of QPO data. In this section we try 
to compare the predictions from the TLM, as indicated by equations in the 
section above, with the observational results.
 
\subsection{Correlation between HBOs and kilohertz QPOs in Z sources}
Recently, Psaltis \etal (1999a) have compiled the low and high-frequency QPO 
data 
of Z sources and atoll sources, as well as black hole candidates. They found a 
tight correlation between HBO frequency and the frequency of lower kHz QPO. The 
data points for the Z sources are consistent with the empirical relation 
$\nu_{HBO}\sim(42\pm3$Hz$)(\nu_1/(500$Hz$))^{0.95\pm0.16} $when the frequency of 
lower 
kHz QPO $\nu_1$ less than 550 Hz. When $\nu_1$ larger than 500 Hz, $\nu_{HBO}$ 
increases slowly with $\nu_1$. This feature seems to be difficult to fit by the RPM. 

 \begin{figure*}[t]
 \psfig{file=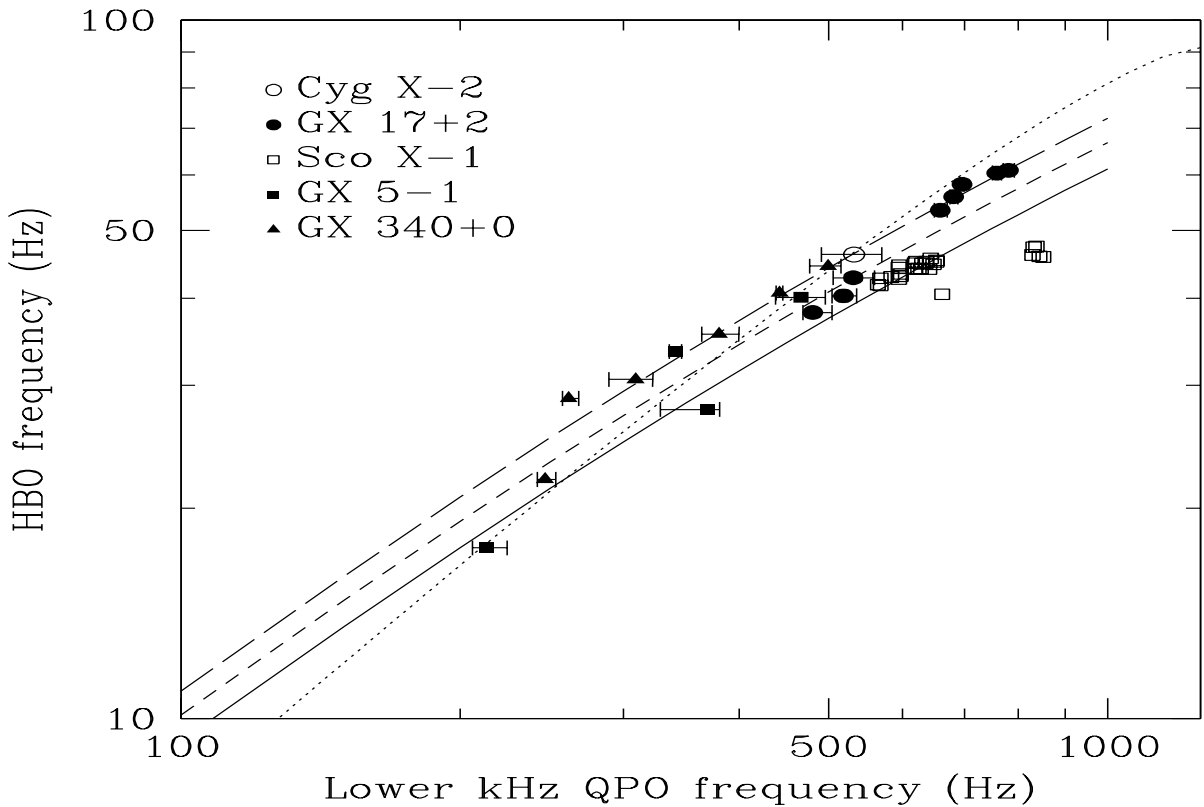,height=5in,width=7in
}
\caption[]{\footnotesize The correlation between HBO frequencies and lower kHz QPO
frequencies of Z sources. The solid, short-dashed and long-dashed
lines
represent TLM predictions with parameters $\Delta\nu=300$ Hz and
$\delta=5.5^o,~6^o$ and $6.5^o$, respectively. The dotted line represents
the prediction of RPM with $M=1.95$\msun~and $a/M=0.22$. The error
bars, if not shown, are comparable to the size of the symbols.
}
\end{figure*}

 \begin{figure*}[t]
 \psfig{file=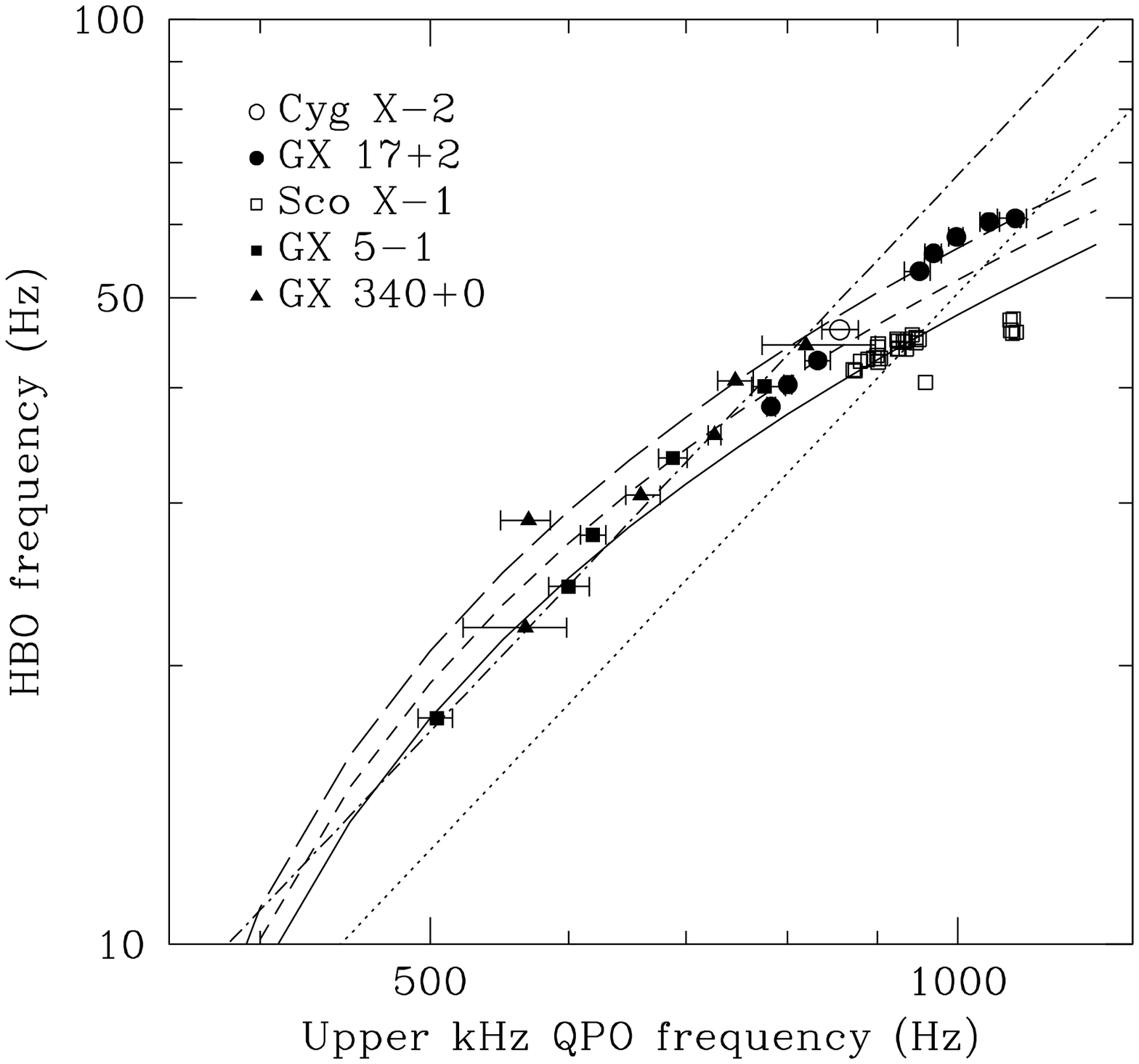,width=7in,height=5in
}
\caption[]{\footnotesize The correlation between HBO frequencies and upper kHz QPO
frequencies of Z sources. The lines have the same meanings as in
Figure1. The dot-dashed line  represents
the prediction of RPM with $M=1.95$\msun~and $a/M=0.30$.
}
\end{figure*}

In 
Figure1, we plot $\nu_{HBO}$ and $\nu_1$ data for five Z sources and compare it 
with the predictions from both RPM and TLM. The fit parameters for RPM 
($M=1.95$\msun, $a/M=0.22$) are taken from Stella \etal (1999) and $a$
is the spin parameter of the neutron star. It is evident 
that RPM predicts a steeper correlation than the observed one and can not explain the 
flatness of such a correlation when $\nu_1>500$ Hz. Any fine tuning of the fit 
parameters $M$ and $a/M$ can not provide a better fit. However, when we plot the 
theoretical curves of TLM according to the equation (5) by choosing quite 
reasonable parameters $\nu_h-\nu_k=300$ Hz and $\delta$ around $6^o$, the match 
of model prediction and observation data is quite well. In addition, we also 
plot the observed correlation between $\nu_{HBO}$ and upper kHz frequency 
$\nu_2$ for five Z sources in Figure 2. In this case, the RPM can not fit the 
observation data if we still choose the parameter $M=1.95$\msun, $a/M=0.22$. 
Even when we choose the fit parameter $M=1.95$\msun, $a/M=0.30$, the fit is 
still not satisfactory. Therefore, our comparison shows that RPM, at least in 
its 
current version, is difficult to fit the correlations between
simultaneously observed HBO
frequencies and 
kHz QPO frequencies. On the contrary, from Figure 1 and Figure 2 we
can see clearly that 
 TLM predicts  rather flatter slopes than
RPM.
 With the 
same parameters, TLM can provide good fits to  both  the $\nu_{HBO}$ -- $\nu_1$ 
and $\nu_{HBO}$ -- $\nu_2$ correlations. 

 For some sources (especially GX17+2 and GX 5-1) in
Figure1, it seems that the slope is steeper than
that predicted by TLM and perhaps matches better with the RPM
prediction. 
However,  we must note that the TLM predictions plotted in
figures 1 and 2
were based on a very simple assumption ($\nu_h - \nu_k = 300$Hz).
In fact,  the peak separation between two kHz QPO frequencies,
$\nu_h-\nu_k$, is certainly not constant for individual sources (see \S
3.4). If we take the observed values of $\nu_h$ and $\nu_k$ directly and use
equation (5) to predict $\nu_L$, the agreement with the observations
will be more close for
individual sources plotted in figures 1 and  2.
 On the other hand, with reasonable parameters, we noted that 
RPM is unable to
fit both the observed correlations shown in figures 1 and 2 
consistently. Even if the RPM prediction could better match the slopes
of  GX17+2 and GX 5-1 in Figure 1, it fails to match the observed
slopes for these sources in Figure 2 if the same parameters were used.
 

\subsection{Difference between low frequency QPOs in atoll sources 
and HBOs in Z sources}
Although there are some reports on the similar QPOs in atoll sources
to HBOs in Z sources (Strohmayer \etal 1996, Wijnands \& van der Klis
1997), it can be clearly seen from Figure 2 in Psaltis \etal (1999a)
that the frequencies of these low-frequency QPOs (10-50 Hz) of atoll
sources have
steeper dependence on the lower kHz QPO frequency than those of HBOs in
Z sources. In Figure 3, we plot this correlation and compare it with
the empirical relation found for HBOs in Z sources. Obviously the
difference between these low-frequency QPOs and HBOs is  large.
It is therefore quite possible that these low-frequency QPOs may have
different origin from HBOs in Z sources. Actually, the similar
low-frequency QPOs (usually seen as extra noise components) have been
found  simultaneously
with HBOs in Z source Sco X-1 and GX 17+2 (van der Klis \etal 1997;
Titarchuk \etal 1999; 
Wijnands \& van der Klis 1999), while the similar HBO-like frequencies have been
identified in two atoll sources 4U 1728-34 and 4U
1702-42 (Titarchuk \etal 1999; Osherovich \& Titarchuk 1999b).

 \begin{figure*}[t]
 \psfig{file=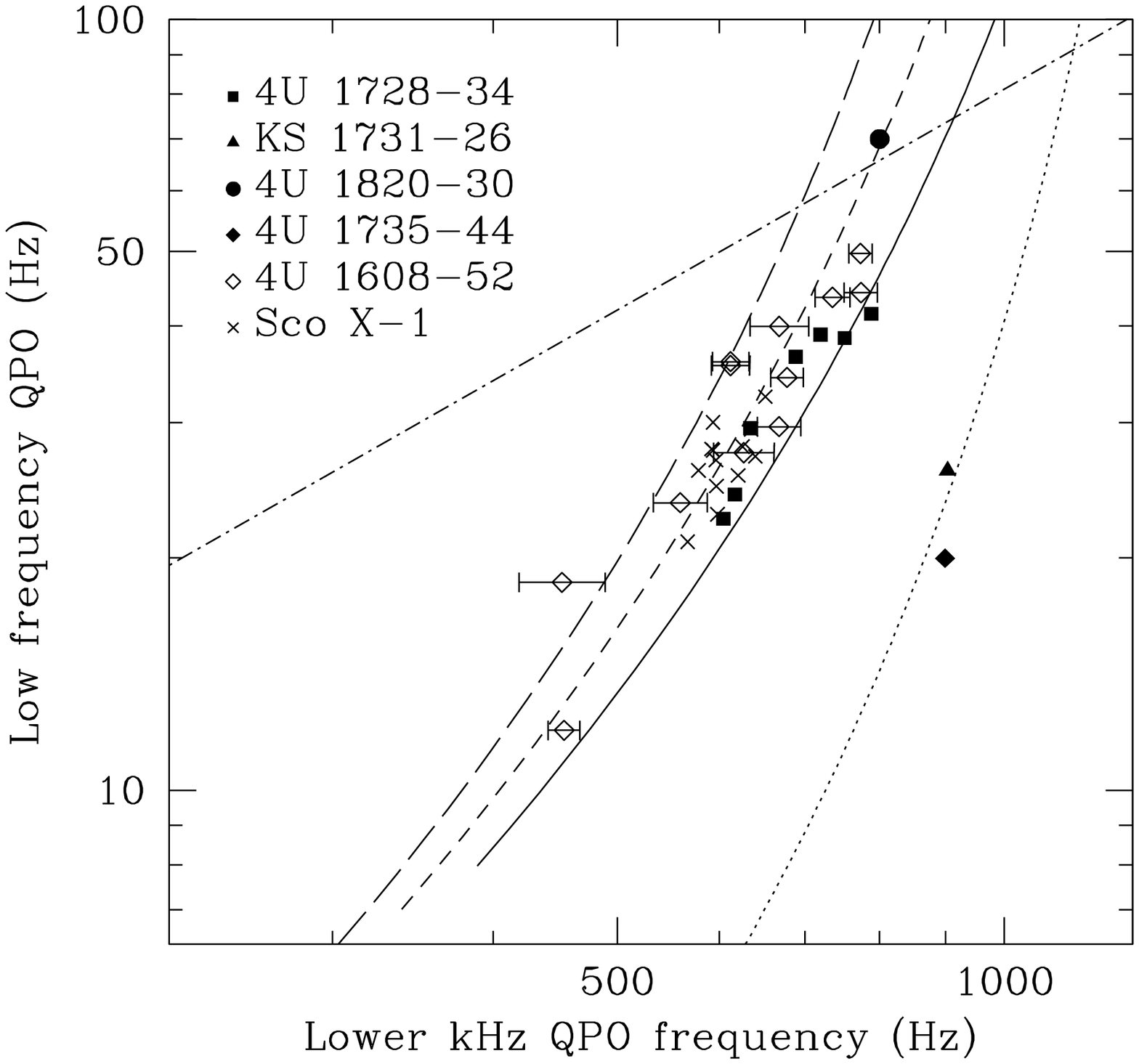,height=5in,width=7in
}
\caption[]{\footnotesize The correlation between  the frequencies of low-frequency
QPOs and lower kHz QPO
frequencies. The solid, short dashed and long dashed lines represent
the TLM predictions with $\alpha (H/R)^2=7.0\times10^{-4}$ and
$M=$1.2\msun, 1.4\msun~and 1.6\msun, respectively. The dotted line
represent the prediction with $\alpha (H/R)^2=3.2\times10^{-4}$ and
$M=$1.4\msun. The dot-dashed line represents the empirical relation
between HBO frequencies and lower kHz QPO frequencies for Z sources.
}
\end{figure*}

In Figure 3, it can be clearly seen that these low-frequency QPOs in
Z source Sco X-1 follow the same correlation with $\nu_1$ as other atoll
sources. This probably suggests that the low-frequency QPOs in both Z
and atoll
sources have the same physics origin and they are obviously different from
HBOs. In
TLM, the frequency of this kind of QPOs is described as the viscous
frequency in 
the
transition layer, therefore the mechanism to produce the low-frequency QPOs is
clearly different from HBOs. Using equations (8) and (9) and assuming
$\alpha (H/R)^2=7.0\times10^{-4}$, $M$ around 1.4\msun~ 
 and
$r_{out}$ in the range of 1.2 to 1.8 (namely $\gamma$ in the range of 6
to 20), 
we plot the
model prediction of TLM in Figure 3 (the result is insensitive to the
spin frequency
of neutron star, $\nu_0=300$ Hz is assumed to  make the plot). We can
see the close agreement
between theory and observations. For KS 1731-26 and 4U
1735-44, the data points are a little away from others. They are
likely associated with the NBOs of Sco X-1 (Psaltis \etal 1999a),
though they can be
still fit by assuming a smaller value of $\alpha(H/R)^2$. However, we
noted that for 4U 1735-44,  recent observations indicated the existence
of a 67Hz QPO together with a possible 900Hz lower kHz QPO (Wijnands
\etal 1998c). If we plot this point in Figure 3, it will also locate in
the range of TLM prediction.

In Figure 3, only the data of February 16, 1996 for 4U 1728-34
are plotted because in these observations the low kHz QPOs can be clearly 
distinguished from the simultaneously observed 
upper kHz QPOs (Ford \& van der Klis 1998). The agreement of the observation
data with our predictions assuming $M$ around 1.4\msun~suggests that the compact
object in 4U 1728-34 may not be a strange star as argued by Li \etal (1999). We 
note that
their conclusion is based on the fit result, $a_k=1.03$,  by Titarchuk \& 
Osherovich (1999), who
used less accurate expression for viscous frequency (see \S 2) and 
ambiguous data for
lower kHz QPO frequencies of 4U 1728-34.

\subsection{Correlation between frequencies of 
low-frequency QPOs and break 
frequencies in Z sources and atoll sources} 
A recent detailed study on the broadband power density spectra of
X-ray binaries indicates that the frequency (1-60 Hz) of low-frequency QPOs
(including HBOs) have a strong correlation with the break frequency
($\nu_{break}$; 0.1-30 Hz) of
both Z sources and atoll sources, possibly as well as black hole
candidates (Wijnands \& van der Klis 1999). However, the HBO frequencies
of Z sources follow a different relationship with $\nu_{break}$ from the
low-frequency QPOs of other sources. Such a difference has not been 
satisfactorily explained so far (see, however,  Titarchuk \etal
(1999) for a possible explanation but note that they mistakenly
plotted the HBO
frequencies of Z sources as the viscous frequencies in their Figure 4). 

 \begin{figure*}[t]
 \psfig{file=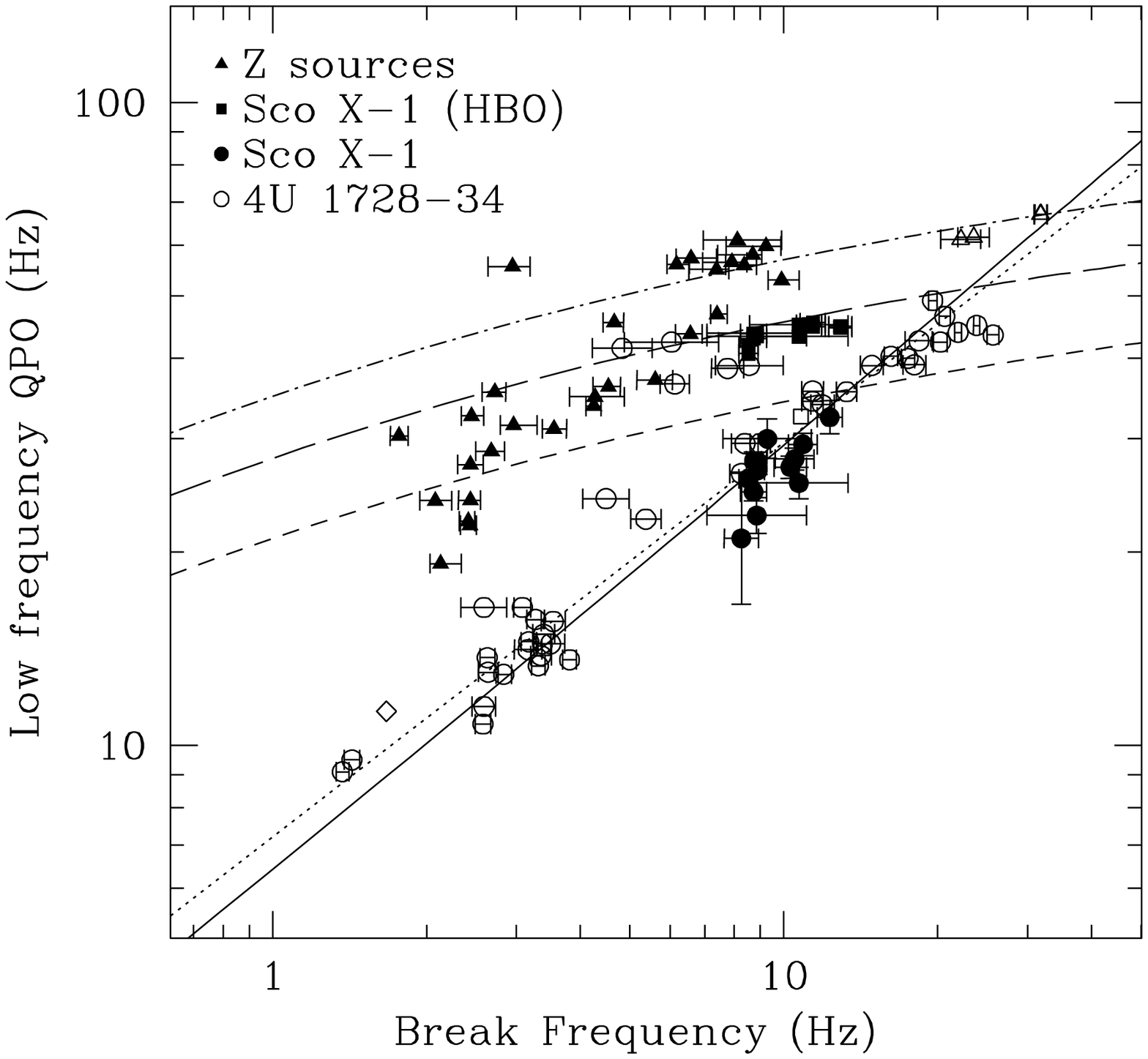,height=5in,width=7in
}
\caption[]{\footnotesize The correlation between  the frequencies of low-frequency
QPOs and break frequencies. The short dashed, long dashed and
dot-dashed lines 
represent
the TLM predictions with $\Delta\nu=300 Hz$ and
$\delta=4.5^o,~6^o, ~7.5^o$, respectively. The solid line
represent the prediction with $\alpha (H/R)^2=7.0\times10^{-4}$ and
$M=$1.4\msun. $l_{d}/R_0=0.3$ was assumed to derive the theoretical
break frequency.
The dotted line represents the empirical relation
$\nu_{break}=0.04\nu_{LF}^{1.63}$ derived for 4U 1728-34. Other atoll
sources are  plotted as open triangles, diamonds and squares.
}
\end{figure*}

In TLM, because HBO and
viscous frequencies come from different mechanisms, it is natural that
they follow different correlations with the break frequencies. According
to equation (5) and equations (8)-(10), we can derive the predicted correlation of
HBO frequencies, viscous frequencies with break frequencies. In Figure 4, we
plot these predictions and compare them with the observation data. For
atoll sources, our prediction closely agrees with the relation,
$\nu_{break}=0.04\nu_{LF}^{1.63}$, which was derived from the
observed correlations $\nu_{LF}$ -- $\nu_{kHz}$ and $\nu_{break}$ -- $\nu_{kHz}$ 
for atoll source 4U 1728-34 (Ford \& van der Klis 1998). The viscous
frequencies of other atoll
sources and Z source Sco X-1 also follow  nearly the same $\nu_{LF}$ --
$\nu_{break}$
correlation. We adopted $l_{d}/R_0=0.3$ in equation (9) and the same parameters 
($\alpha
(H/R)^2=7.0\times10^{-4}$, M=1.4\msun) as above to derive this
correlation. The value, $l_{d}/R_0=0.3$, is appreciated by the
observed data and  also consistent with our assumption that the diffusion
length scale
is smaller than the vertical and radial size of the transition layer. 
 Note that $l_d = 0.3 R_0 < H $ means $H/R$ may be larger than
0.3. This is quite possible if the magnetic field is low and the radial size of transition layer is
small.
If we take $\alpha
(H/R)^2=7.0\times10^{-4}$, it requires that $\alpha$ should be less than
0.01. Such a lower value of $\alpha$ is still well within the
range of viscosity parameter discussed in the accretion disk model with
boundary layer.
Assuming
$\nu_h-\nu_k=300$ Hz and $\delta$ around $6^o$, we derive the
correlation between HBO frequencies and break frequencies. Comparing with the
observation data of Z sources, we note that the agreement is
well. There are several data points for atoll source 4U 1728-34
locating at the branch for HBO in Z sources, which may indicate that
these data are actually frequencies of HBO-type QPOs in atoll
sources (see \S 3.2). 

\subsection{Variation of peak separation of two kilohertz QPOs}

The observed variation of difference of two kHz QPO frequencies in at 
least one Z source, Sco X-1 (van der Klis \etal 1997) and four atoll sources, 
4U 1728-34, 4U1608-52, 4U 1735-44 and 4U 1702-43 (M\'endez \etal 1998a,b; Ford 
\etal 1998; M\'endez \& van der Klis 1999;
Markwardt \etal 1999), seem to exclude
any QPO model that predicts the constant peak separation. The magnetospheric
beat-frequency model, as well as the sonic-point beat-frequency model, 
probably can not be responsible for the kHz QPOs, unless further
modifications are made.

\begin{figure*}[t]
 \psfig{file=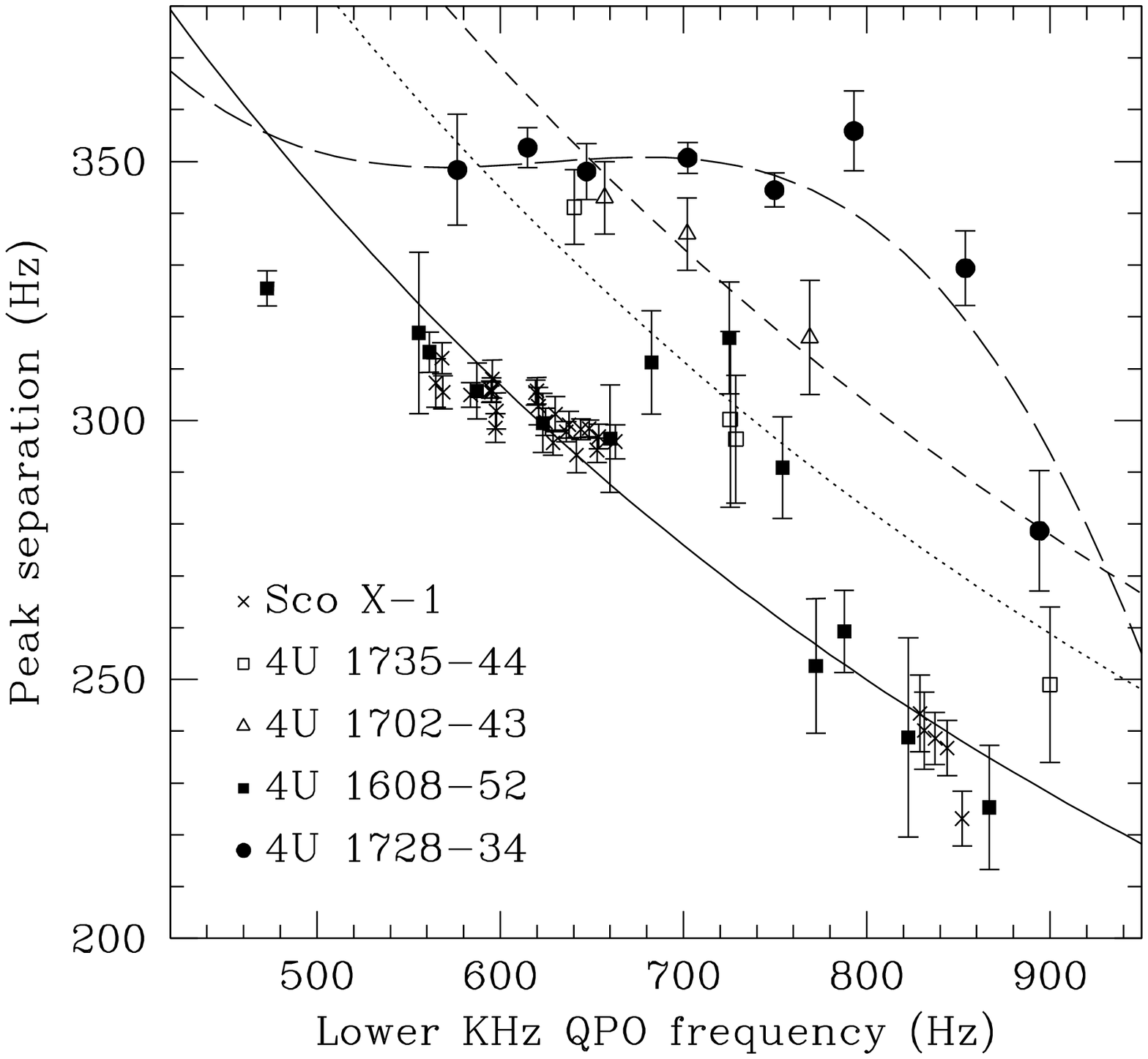,height=5in,width=7in
}
\caption[]{\footnotesize The relation between  the peak separations of two kHz QPO
frequencies
and the lower kHz QPO frequencies. The solid, dotted and short dashed
lines represent
the TLM predictions using equation (6) with $\Omega_0/2\pi=$340 Hz, 365 Hz and 
380 Hz,
respectively. The long dashed line represent the prediction using
equations (3) and (6) with parameters $\nu_0=405$ Hz, $c_1=0.26$ Hz$^{-1/6}$ and 
$
c_2=3.4\times10^{-5}$ Hz$^{-3/2}$.
}
\end{figure*}

Recent precise measurements of kHz QPO frequencies show that the peak
separation $\Delta\nu=\nu_2-\nu_1$ decreases considerably when kHz QPO 
frequencies 
increase (van der Klis 2000). This feature can not be well fitted by the RPM, 
which predicts a much 
steeper decrease of $\Delta\nu$ when $\nu_2$ increases to more than 1000 Hz and 
a significant decrease of
$\Delta\nu$ when $\nu_2$ decreases to less than 700 Hz (Stella \& Vietri 1999). 
The latter 
effect has not been observed yet. A precise match between RPM and observations 
requires some additional assumptions such as the dependence of orbital 
eccentricity on
the orbital frequency. In TLM, however, the decrease of $\Delta\nu$ as the 
increase of kHz QPO
frequency is a natural result even if we simply take $\Omega=constant$ (see equation (6)). In Figure 
5, we show the $\Delta\nu
-\nu_1$ relation for all 5 sources with observed $\Delta\nu$ variations. When 
we plot the TLM prediction
assuming the constant $\Omega/2\pi$ (340Hz, 365Hz and 380Hz) in equation (6), we see 
that 
even at this simple assumption the theoretical predictions
are closely agreement with most observation data  of Sco X-1, 
4U 1708-52, 4U 1735-44 and 4U 1702-43.
However, it is clear that this simple assumption can not explain the data of 
4U 1728-34. In fact, the matter in magnetosphere probably experiences
differential rotation and $\Omega$ may not be a 
constant. If we assume that the angular velocity could be expressed as
$
\frac{\Omega}{2\pi}=\nu_0-(c_1\nu_k^{2/3}-c_2\nu_k^{2})^2
$ (Osherovich \& Titarchuk 1999a)
where $\nu_0$, $c_1$ and $c_2$ are constants,
and take $\nu_0=405$ Hz, $c_1=0.26 $Hz$^{-1/6}$ and $
c_2=3.4\times10^{-5} $Hz$^{-3/2}$, we can see the prediction
 of peak variation of twin kHz QPO frequencies is close to the
observation data for 4U 1728-34.
In summary, it seems that  the variation of 
peak 
separation of two kHz QPOs is also consistent with the theoretical
prediction
of TLM.

\section{Observational determination of misalignment of neutron star 
magnetosphere and accretion disk}

There are only few methods  that can be used to estimate the
magnetic inclination angle between magnetic axis and neutron star
rotation axis for radio pulsars (Lyne \& Manchester 1988). It may be more
difficult
to measure these angles for X-ray pulsars and low-mass X-ray binaries. However,
according to TLM (see equation (7)), if we have simultaneous
observation data for frequencies of both kHz QPOs and HBOs, we may
estimate the angle between magnetophere equator and disk plane for
neutron star X-ray binaries. We note that  equation (7) is derived under the assumption
that $\delta$ is small. In fact, a more accurate equation, which is
also applicable
to arbitrary values of $\delta$, can be
directly
obtained from the dispersion relation of Keplerian oscillator 
(see equation (2)), namely
\begin{equation}
\delta=arcsin(\frac{\nu_2\nu_{HBO}}{\nu_1\sqrt{\nu_2^2+\nu_{HBO}^2-\nu_1^2}}),
\end{equation}
where $\nu_1$, $\nu_2$ and $\nu_{HBO}$ are the observed lower, upper kHz QPO
and HBO frequencies respectively, and we assume that they correspond
to the Keplerian frequency and two characteristic oscillation
frequencies derived from the dispersion relation (see \S 2). In the limit
$\nu_{HBO}<<\nu_2$, equation (11) is identical to equation (7).
Note that equation (11) can be also used as a test of TLM because
the derived values of $\delta$  using different set of simultaneously
 observed QPO frequency data 
for each source should be nearly the same (Titarchuk \& Osherovich 2000).

\begin{table*}
\caption{Angle  between magnetosphere equator and accretion disk plane
determined by simultaneously observed QPO frequency data}
\begin{center}
\begin{tabular}{lcccclc}
\\ \hline
Name	&Type	&$\nu_2$	&$\nu_1$ &	$\nu_{HBO}$& 	Ref.	&$<\delta>$\\
 & &(Hz)&(Hz)&(Hz)& & ($ ^o$)\\
\hline
Cyg X-2		&Z		&   856		&	532	& 46	&1,2
	&6.34$\pm$0.14\\
GX 17+2		&Z		&   640--1090	&	480--780	&25--61
	&1,3	&6.30$\pm$0.37\\
Sco X-1		&Z		 &  870--1080		&550--860	&41-48
	&1,4	&5.44$\pm$0.22\\	
GX 5-1		&Z		&500--890		&	210--660	&17--51
	&1,5	&5.80$\pm$0.62\\
GX 340+0	&Z		&560--820		&	250--500&22-45	&1,6
	&6.40$\pm$0.45\\
4U 1728-34	&Atoll		&930--1130		&600--790	&52-90	&1,7,8
	&8.01$\pm$1.06\\
4U 1702-42	&Atoll		&1000--1080&660-770	&33--40	&9,10
	&3.96$\pm$0.23\\
\hline
\end{tabular}
\\ \vskip 1mm
\end{center}
\noindent
References: (1). Psaltis et al. (1999a); (2). Wijnands et al. 1998a;
(3). Wijnands et al. 1997; (4) van der Klis et al. 1997;
(5). Wijnands et al. 1998b; (6). Jonker et al. 1998; (7). Strohmayer et
al 1996; (8). Ford \& van der Klis 1998; (9). Markwardt et al. (1999);
(10). Osherovich \& Titarchuk (1999b)
\end{table*}

Up to know, simultaneous frequency data of both kHz QPOs and 
HBOs  have been
obtained with RXTE for five Z sources. In two atoll sources with
simultaneous data of two KHz QPO frequencies, the HBO-like
frequencies have been
identified recently (Titarchuk \etal 1999; Osherovich \&
Titarchuk 1999b). Therefore, the misalignment of magnetosphere and
accretion disk can be estimated for these sources according to
equation (11). In table 1, we list the
observation data and the derived average values of $\delta$ for them.
It is clear that all of them have smaller $\delta$ values and are
therefore consistent with the assumption in \S 2. Especially for
Z sources, their $\delta$ values are in a very narrow range from $5.^o4$
to $6.^o4$.
Moreover,  the standard derivations of derived $\delta$
values are small, which means nearly the same $\delta$ value is
obtained for each source  using the simultaneously observed QPO
frequency data. This is well agreement with the prediction of TLM.
In addition,
we noted that the method we used
to determine $\delta$  is better than that used by Osherovich \&
Titarchuk (1999b) and Titarchuk \etal (1999), who derived $\delta$ by
fitting $\Omega$ calculated with the observed kHz QPO frequencies (see
equation (2)). It is obviously not necessary to do so if we have
simultaneously observed frequency data of HBOs and kHz
QPOs. Instead, it
is more straightforward and accurate to derive $\delta$ by using equation (11).
However, their estimated values, $\delta=3.^o9\pm0.^o2$ for 4U 1702-42,
$\delta=8.^o3\pm1.^o0$ for 4U 1728-34 and $\delta=5.^o5\pm0.^o5$ for
Sco X-1, are very close to our estimations. For other four Z sources in
Table 1, it is the first time to give the estimations of their angles
between magnetosphere equator and disk plane.

\section{Discussion}

The comparisons of the model predictions of transition layer model with the 
observed
data seem to suggest that TLM
may provide better explanations on them than
the current version of other QPO models. In contrary to the inability of beat-frequency model to explain 
the 
variation of peak separation of two kHz QPO frequencies and the difficulty of 
relativistic precession model to explain both the correlations of HBO
frequencies with lower and upper kHz
QPO frequencies, TLM is so far the only model that can
self-consistently explain  these observed
correlations.  In addition, TLM suggests some physics mechanisms for
producing low-frequency QPOs 
and break frequencies in X-ray binaries, and can explain 
the observed correlation between the frequencies of low-frequency QPOs
and the break frequencies. On the contrary, no clear mechanisms for these low frequency
features have been discussed yet in most of other QPO models.

Another clear difference between TLM and other models is that the
lower kHz QPO frequency, rather than the upper kHz QPO frequency, is
described as Keplerian frequency. Therefore, it is not necessary in
TLM to involve larger neutron star mass and stiff equation of state as
required in other models such as sonic-point beat-frequency model
(Miller \etal 1998) and relativistic precession model
(Stella \& Vietri 1998, 1999).
The compact star in TLM can be still a normal neutron star with mass 
around 1.4\msun~and radius
around 12km. Indeed, the observed correlation of QPO frequencies
 can be well explained by taking these parameters in TLM. A larger neutron
star mass ($>$ 2\msun) is not appreciated since it will lead to
steeper correlation of low-frequency QPO frequencies  
with lower kHz QPO frequencies than that observed for atoll sources 
(see Figure 2).

The variations of the size of magnetosphere and the size of transition layer
due to the change of mass accretion rate may account for the the observed
variations of QPO frequencies in X-ray binaries. Because the size of 
magnetosphere can be
described approximately by the Alv\'en radius $R_A$, which is inverse
proportional to $\dot{M}^{-2/7}$,
the magnetosphere will shrink if accretion rate $\dot{M}$ increases. This may lead to
the increases of the kHz QPO frequencies ($\nu_h$, $\nu_k$) and the HBO frequency 
($\nu_L$). The size of the transition layer can be described by
$R_{out}-R_0$ , which is directly
related to the Reynolds number of accretion flow (see equation
(9)). The increase of  accretion
rate will cause the increase of Reynolds number, and therefore lead to
the decrease of $R_{out}-R_0$.
According to equations (8) and (10), both the viscous frequency and
the break frequency will
increase as the increase of accretion rate. Indeed, many observations
have shown the increases
of  kHz QPO frequencies, HBO frequency and break frequency when the sources move
along the tracks in the color-color diagram, equivalently, in the
sense of increasing inferred
accretion rate (e.g. M\'endez \& van der Klis 1999). These results are
also consistent with the predictions of TLM. Recently, Cui (2000)
proposed that the disappearance of kHz QPOs may be explained by the 
``disengagement'' between the magnetosphere and the Keplerian
disk. Campana (2000) suggested that the vanishing of the magnetosphere
may led to the stopping of the kHz QPO activity. This could be also
the natural result of TLM since the Keplerian oscillator
has no chance to enter the rotating frame of reference at all if it
losts direct contact with the magnetosphere.
 
We note that the similar QPO frequencies in the range of 1--400 Hz
have been observed in X-ray millisecond pulsar SAX J1808.4-3658 
(Wijnands \& van der Klis 1998a) and Galactic black hole candidates
GRO J1655-40 (Remillard \etal 1999a), GRS 1915+105 (Morgan, Remillard
\& Greiner 1997), XTE J1550-564 (Remillard \etal 1999b) and XTE
J1859+226 (Cui \etal 2000).  It is
still not clear whether the 67--400
Hz QPOs observed in these objects are identical to the kHz QPOs in
neutron star LMXBs. But if we
assume that these hundred-hertz QPOs are similar to the lower kHz QPOs,
the correlations of their  frequencies of low-frequency QPOs (1-20Hz)
and break frequencies with the frequencies of these `kHz QPOs' seem to
be   similar  as found for atoll sources (see  Psaltis \etal
1999a; Wijnands \& van der Klis 1999).  In fact, if we plot the QPO
frequency data for  SAX J1808.4-3658 in Figures 3 and 4, these
points will locate in the lowerleft parts of both figures  but are
still consistent with the correlations for atoll sources. Their
positions in these two figures also support that
SAX J1808.4-3658 is similar to the low-luminosity LMXBs with low
accretion rate (Wijnands \& van der Klis 1998a).
Its estimated magnetic field strength, $B<(2-6)\times 10^8$ Gauss (Wijnands \& van
der Klis 1998b), is indeed similar as that for atoll sources. 
For black hole candidates, a similar transition layer as in LMXBs may
also exist between a geometrically thin clod disk and a geometrically
thick hot disk (or accretion-dominated accretion flow; see Narayan \& Yi 1994) due to the sub-Keplerian rotating
feature of the inner hot disk. Because of the lack of magnetosphere
near the black hole, we can only predict 
three characteristic frequencies ($\nu_k$, $\nu_v$ and $\nu_{break}$)
from TLM. This may help us to understand
why we did not observed QPO frequencies larger than 400 Hz for black
hole candidates and why their power density spectra and correlations
between low and high QPO
frequencies are more similar as those for atoll sources than  for Z
sources (van der Klis 1994, Psaltis \etal 1999a). A detailed
comparison of these features, however,  is beyond the scope of this paper. 

We should also mention that  as in other QPO models,
TLM requires blob be lifted from the inner part of Keplerian disk. It is
still unclear which mechanism could lead to such lifting. In
addition, most theoretical predictions of TLM are made by assuming the
angular velocity of magnetosphere $\Omega$ is constant. This may not be the
case in reality. Considering a non-constant  $\Omega$ will lead to the 
complexity of the dispersion relation and therefore some different
characteristic oscillation frequencies. Moreover, the lack of explicit
knowledge about the diffusion process in the transition layer may
result in  the uncertainty in estimating the break frequency. In spite
of these questions, TLM should be considered  a very
competitive model
for QPOs in X-ray binaries.

\acknowledgements

I am very grateful to Dimitrios Psaltis, Eric Ford, Michiel van der Klis,
Rudy Wijnands, Mariano M\'endez and Lev Titarchuk for 
kindly providing me the data files. I thank Wei Cui   and the
anonymous referee for  helpful comments, and thank Lev Titarchuk, Wenfei Yu and 
Chengmin Zhang for stimulating discussions.


\begin{references}  

\reference{akhiezer}
Akhiezer, A.I., Akhiezer, I.A., Polovin, R.V., Sitenko, A.G., Stepanov, K.N., 
1975,
Plasma Electrodynamics (Oxford: Pergamon)
\reference{alpar}
Alpar, M.A., Shaham, J., 1985, Nature, 316, 239
\reference{benson}
Benson, R.F., 1977, Radio Sci., 12, 861
\reference{campana2000}
Campana, S., 2000, ApJ, 534, L79
\reference{cui2000}
Cui, W., 2000, ApJ,  534, L31
\reference{cui2000b}
Cui, W., Shrader C.R., Haswell, C.A., Hynes, R.I., 2000, ApJ, 535, L123
\reference{finn}
Finn, L.S., 1994, Phys. Rev. Lett., 73, 1878
\reference{fordk}
Ford, E.C., van der Klis, M., 1998, ApJ, 506, L39
\reference{ford}
Ford, E.C., van der Klis, M., van Paradijs, J., M\'endez, M., Wijnands, R., 
Kaaret, P., 1998, ApJ, 508, L155
\reference{hasinger}
Hasinger, G., van der Klis, M., 1989, A\&A, 225, 79
\reference{jonker}
Jonker, P., Wijnands, R., van der Klis, M., Psaltis, D., kuulkers,
E., Lamb, F.K., ApJ, 1998, 499, L191
\reference{lai}
Lai, D., Lovelace, R., Wasserman, I., 1999, ApJ, submitted (astro-ph/9904111)
\reference{landau}
Landau, L., Lifshitz, E., 1960, Mechanics (New York: Pergamon)
\reference{li}
Li, X.D., Ray, S., Dey, J., Dey, M., Bombaci, I., 1999, ApJ, 527, L51
\reference{lyne}
Lyne, A.G., Manchester, R.N., 1988, MNRAS, 234, 477
\reference{markwardt}
Markwardt, C.B., Strohmayer, T.E., Swank, J.H., 1999, ApJ, 512, L125
\reference{mendez}
M\'endez, M., van der Klis, M.,1999, ApJ, 517, L51
\reference{mendez}
M\'endez, M., van der Klis, M., van Paradijs, J., Lewin, W.H.G., Vaughan, B.A. 
\etal 1998a, ApJ, 494, L65
\reference{mendez}
M\'endez, M., van der Klis, M., Wijnands, R., Ford, E.C., van
Paradijs, J.,  Vaughan, B.A., 1998b, ApJ, 505, L23
\reference{middleditch}
  Middleditch, J., Priedhorsky, W.C., 1996, ApJ, 306, 230
\reference{miller}
Miller, M.C., Lamb, F.K., Psaltis, D., 1998, ApJ, 508, 791
\reference{morgan}
Morgan, E.H., Remillard, R.A., Greiner, J., 1997, ApJ, 482, 993
\reference{narayan}
Narayan, R., Yi, I., 1994, ApJ, 428, L13
\reference{osherovich}
Osherovich, V., Titarchuk, L., 1999a, ApJ, 522, L113
\reference{osherovich}
Osherovich, V., Titarchuk, L., 1999b, ApJ, 523, L73
\reference{popham}
Popham, R., Sunyaev, R., 2001, ApJ, 547, 355
\reference{psaltis}
Psaltis,D., M\'endez, M., Wijnands, R., Homan, J., Jonker, P.G., \etal 1998, 
ApJ, 501, L95
\reference{psaltis}
Psaltis,D., Belloni, T., van der Klis, 1999a, ApJ, 520, 262
\reference{psaltis}
Psaltis, D., Norman, C., 2000, ApJ, submitted (astro-ph/0001391)
\reference{psaltis}
Psaltis, D., Wijnands, R., Homan, J., Jonker, P.G., van der Klis, M.,
\etal 1999b, ApJ, 520, 763
\reference{remillard}
Remillard, R.A., McClintock, J.E., Sobczak, G., Bailyn, C.D., Orosz,
J.A., 1999b, ApJ, 517, L127
\reference{remillard}
Remillard, R.A., Morgan, E.H., McClintock, J.E., Bailyn, C.D., Orosz,
J.A., 1999a, ApJ, 522, 397
\reference{shakura}
Shakura, N.I., Sunyaev, R.A., 1973, A\&A, 24, 337
\reference{stella}
Stella, L., Vietri, M., 1998, ApJ, 492, L59
\reference{stella}
Stella, L. Vietri, M., 1999, Phys. Rev. Lett., 82, 17
\reference{stella}
Stella, L., Vietri, M., Morskink, S.M., 1999, ApJ, 524, L63
\reference{strohmayer}
 Strohmayer, T.E., Zhang, W., Swank, J.H., Smale, A., Titarchuk, L.,
 Day, C., 1996, ApJ, 469, L9
\reference{thorsett}
Thorsett, S.E., Chakrabarty, D., 1999, ApJ, 512, 288
\reference{titarchuk}
Titarchuk, L., Osherovich, V., Kuznetov, S., 1999, ApJ, 525, L129
\reference{titarchuko}
Titarchuk, L., Osherovich, V., 1999, ApJ, 518, L95
\reference{titarchuko00}
Titarchuk, L., Osherovich, V., 2000, ApJ, 537, L39
\reference{vdklis}
van der Klis, M., 1994, A\&A, 283, 469
\reference{vdklis}
van der Klis, M., 2000, ARA\&A, 38, 71
\reference{vdklis}
  van der Klis, M., Jasen, F., van Paradijs, J., Lewin, W.H.G., van den Heuvel,
E.P.J., 1985, Nature, 316, 225
\reference{vdklis}
 van der Klis, M., Swank, J.H., Zhang, W., Jahoda, K., Morgan, E.H., 
\etal 1996, ApJ, 469, L1
\reference{vdklis}
van der Klis, M., Wijnands, R., Horne, K., Chen, W., 1997, ApJ, 481, L97 
\reference{wijnandsh97}
 Wijnands, R. Homan, J., van der Klis, M., Kuulkers, E., van Paradijs, J., Lewin, W.H.G.,
          Lamb, F.K., Psaltis, D., Vaughan, B. 1997, ApJ, 490, L157
\reference{wijnandsh98}
 Wijnands, R. Homan, J., van der Klis, M., Kuulkers, E., van Paradijs, J., Lewin, W.H.G.,
          Lamb, F.K., Psaltis, D., Vaughan, B. 1998a, ApJ, 490, L87
\reference{wijnands}
Wijnands, R., M\'endez, M., van der Klis, M.,  Psaltis, D., kuulkers,
E., Lamb, F.K., ApJ, 1998b, 504, L35
\reference{wijnandsk}
Wijnands, R., van der Klis, M., 1997, ApJ, 482, L65
\reference{wijnandska}
Wijnands, R., van der Klis, M., 1998a, ApJ, 507, L63
\reference{wijnandskb}
Wijnands, R., van der Klis, M., 1998b, Nature, 394, 344
\reference{wijnands}
Wijnands, R., van der Klis, M., 1999, ApJ, 514, 939
\reference{wijnands}
Wijnands, R., van der Klis, M., M\'endez, M., van Paradijs, J., Lewin,
W.H.G., Lamb, F.K., Vaughan, B., ApJ, 1998c, 495, L39


\end{references}
\end{document}